\documentclass[12pt]{article}
\usepackage{amsmath}
\usepackage{amsthm}
\usepackage{graphicx,psfrag,epsf}
\usepackage{enumerate}
\usepackage{natbib}
\usepackage{url} 
\usepackage{ragged2e}
\usepackage{subfig}
\usepackage{float}
\usepackage{adjustbox}
\usepackage{framed}

\newcommand{\blind}{0}

\DeclareMathOperator{\bX}{\boldsymbol{X}}
\DeclareMathOperator{\bI}{\boldsymbol{I}}
\DeclareMathOperator{\bone}{\boldsymbol{1}}

\DeclareMathOperator{\bx}{\boldsymbol{x}}
\DeclareMathOperator{\ba}{\boldsymbol{a}}
\DeclareMathOperator{\bb}{\boldsymbol{b}}
\DeclareMathOperator{\bZ}{\boldsymbol{Z}}
\DeclareMathOperator{\bz}{\boldsymbol{z}}
\DeclareMathOperator{\bV}{\boldsymbol{V}}
\DeclareMathOperator{\bv}{\boldsymbol{v}}
\DeclareMathOperator{\be}{\boldsymbol{e}}
\DeclareMathOperator{\bD}{\boldsymbol{D}}
\DeclareMathOperator{\bL}{\boldsymbol{\Lambda}}
\DeclareMathOperator{\bU}{\boldsymbol{U}}
\DeclareMathOperator{\bu}{\boldsymbol{u}}
\DeclareMathOperator{\b0}{\boldsymbol{0}}

\begin{document}

	\def\spacingset#1{\renewcommand{\baselinestretch}%
		{#1}\small\normalsize} \spacingset{1}

	
	\if0\blind
	{
		\title{\bf Generalized Biplots for Multidimensionally Scaled Projections}
		\author{J.T. Fry, Matt Slifko, and Scotland Leman\\
			Department of Statistics, Virginia Tech}
		\maketitle
	} \fi
	
	\if1\blind
	{
		\bigskip
		\bigskip
		\bigskip
		\begin{center}
			{\LARGE\bf Generalized Biplots for Multidimensionally Scaled Projections}
		\end{center}
		\medskip
	} \fi
	
	\bigskip
	\begin{abstract}
		Dimension reduction and visualization is a staple of data analytics. Methods such as Principal Component Analysis (PCA) and Multidimensional Scaling (MDS) provide low dimensional (LD) projections of high dimensional (HD) data while preserving an HD relationship between observations. Traditional biplots assign meaning to the LD space of a PCA projection by displaying LD axes for the attributes. These axes, however, are specific to the linear projection used in PCA. MDS projections, which allow for arbitrary stress and dissimilarity functions, require special care when labeling the LD space. We propose an iterative scheme to plot an LD axis for each attribute based on the user-specified stress and dissimilarity metrics. We discuss the details of our general biplot methodology, its relationship with PCA-derived biplots, and provide examples using real data.
	\end{abstract}
	
	\noindent%
	\vfill
	
	\newpage
	\spacingset{1.45} 
	\section{Introduction}
	\label{sec:intro}
	
	Dimension reduction and data visualization are staples of any good analysis, whether as an exploratory or disseminating tool. Visualizations provide an opportunity for the analyst to discover underlying structures and gain insights not easily gleaned by examining the raw data itself \citep{keim2002information}. Techniques range from the simplistic and easily interpretable, such as univariate histograms and bivariate scatterplots, to more complicated dimension reduction procedures, such as Multidimensional Scaling (MDS) and Principal Component Analysis (PCA). Particularly, as the dimensionality of data increases, most bivariate visualizations fail to capture all of the intricacies contained within the data. It is common to use lower dimensional approximations of a data structure in order to gain some understanding of the complexity involved.
	
\begin{figure}[H]
		\centering
		\includegraphics[width=5in]{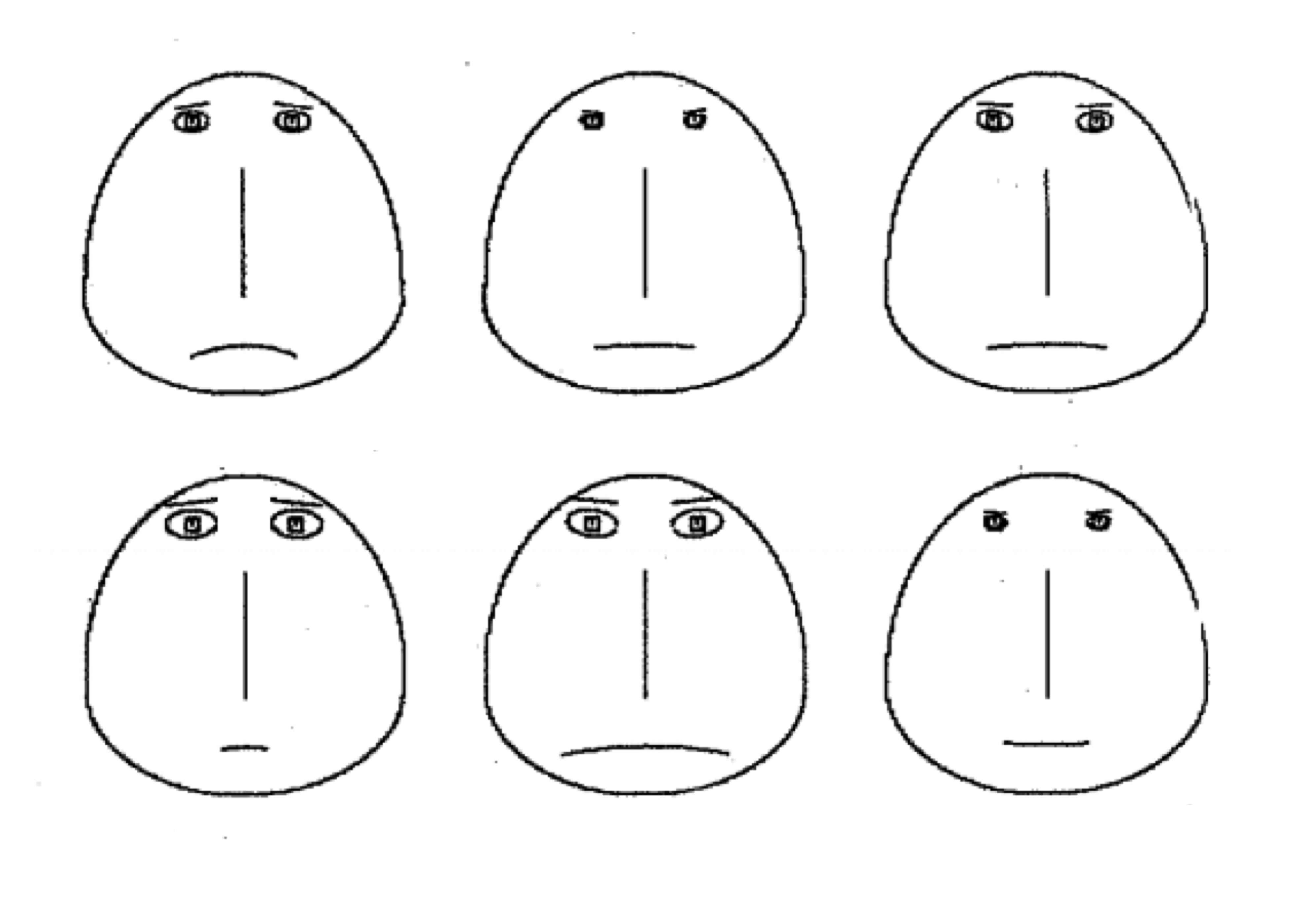}
		\caption{Example of Chernoff's faces \citep{chernoff1973use}.  Facial features (e.g. eyes, nose, mouth) for each face are representative of the high dimensional attributes.} \label{chernoff_figure}
	\end{figure}

\begin{figure}[H]
		\centering
		\includegraphics[width=5in]{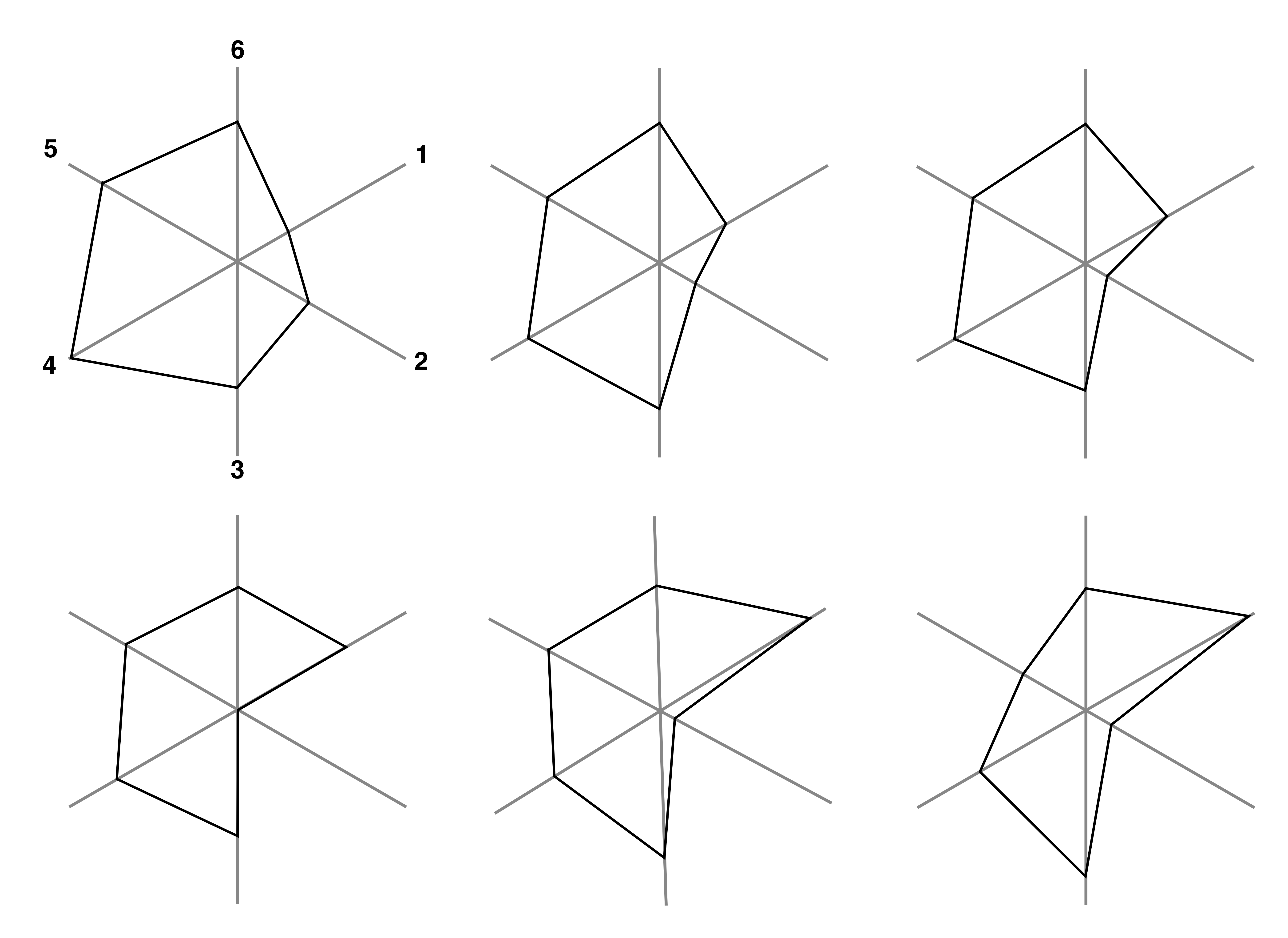}
		\caption{Example of a starplot with $6$ attributes ($1-6$). Plots with similar shapes have similar high dimensional attributes.} \label{starplot_figure}
	\end{figure}

	Several visualization techniques attempt to display all the high dimensional attributes for each observation using a single plot. One of the earliest and most well known attempts is Chernoff's faces \citep{chernoff1973use}, where each observation is represented by different characteristics on a face, such as the length of the nose or the curvature of the mouth. Similar-looking faces can be to grouped to represent similarities in the high dimensional data (Figure \ref{chernoff_figure}).  Unfortunately, this technique is limited to $18$ attributes and the decision of which variables are assigned to which facial features can impact the conclusions. Star plots \citep{chambers1983graphical} provide improvements on both of these limitations. A star plot consists of equiangular spokes, one for each attribute, emanating from a central point. The length of one spoke represents the value of the attribute for a particular observation relative to the maximum value across all observations (Figure \ref{starplot_figure}). Much like the Chernoff faces, each observation produces its own figure, with similar shapes being grouped together. In applications that utilize a large number of attributes cannot be reasonably displayed using the previously mentioned techniques. 
	
	Among the most popular techniques to reduce dimensionality are the aforementioned PCA and MDS methods. PCA provides a new set of orthogonal axes in the directions that maximize the variance of the reduced dimensional, projected data. To produce the low dimensional projection, the user removes the axes that capture the lowest amounts of variance. MDS provides additional flexibilities by allowing users to interchange dissimilarity metrics in both the high dimensional attribute space and the projected space.  MDS algorithms project the data such that, on average, the low dimensional dissimilarities most closely matches the high dimensional dissimilarity. It should be mentioned that PCA is a specific case of an MDS algorithm, which we provide details in section \ref{sec:classical_mds}.
	
	Although MDS preserves the average dissimilarity between observations, we lose a sense of how the original attributes affect positioning. For example, the PCA axes are principal components representing a linear combination of the attributes. To rectify this, researchers have developed ways of labeling the low dimensional space. Gabriel \citep{gabriel1971biplot} developed the original biplot, a PCA-specific technique that adds vectors to the PCA projection to represent a projection of the high dimensional axes. \cite{cheng2016data} propose the Data Context Map, which displays both the observations and attributes as points in the same space. This is achieved by creating a large composite matrix with observations and attributes that are treated as observations. As a consequence, the projection of the observations is affected by the treatment of the attributes, instead of simply labeling the already created projection. Gower \citep{gower1992generalized} expanded upon the PCA biplot by allowing other distance metrics. Using an approximation based on Euclidean distance, axes are linearly projected based on the specified distance function. Referred to as the nonlinear biplot, these projections often create highly curved low dimensional axes.
	
	The remainder of the manuscript is organized as followed. First, we review PCA and MDS, while establishing the connection between them. We then discuss the PCA biplot, introduce our method for the generalized MDS biplot, and show the association between both techniques. Finally, we apply our generalized MDS biplot to a real dataset and discuss the generated projections.
	
	\section{Notation}
	For clarity and ease of reading, we define some notation that will be used throughout the manuscript. We let $\bX=(\bx_1,\ldots,\bx_n)'$ denote an $n \times p $ matrix of high dimensional data containing $n$ observations of $p$ continuous attributes. We assume that $\bX$ is full column rank and has unitless columns with column means of $0$. Utilizing the singular value decomposition (SVD), we can write $\bX=\bU\bL^{1/2}\bV'$, where  $\bL=diag(\lambda_1,\ldots,\lambda_p)$ ( $\lambda_1\geq\lambda_2\geq\ldots\geq\lambda_p$) is a diagonal matrix with the $p$ positive eigenvalues of $\bX'\bX$ and $\bX\bX'$ in descending order, and $\bU = (\bu_1,\ldots,\bu_p)$ and $\bV=(\bv_1,\ldots,\bv_p)$ are  $n \times p$  and $p \times p$ orthonormal matrices whose columns contain the eigenvectors of $\bX\bX'$ and $\bX\bX'$, respectively. We can further partition the SVD as $\bX = (\bU_1,\bU_2)diag(\bL_1,\bL_2)^{1/2}(\bV_1,\bV_2)'$ where $\bL_1$ contains the first $m$ eigenvalues and $\bU_1$ and $\bV_1$ contain the corresponding eigenvectors. We let  $\bZ=(\bz_1,\ldots,\bz_n)'$ be an $n \times m$, $m<p$ matrix of low dimensional coordinates corresponding to $\bX$. Similarly to $\bX$, $\bZ$ can be decomposed into  $\tilde{\bU}\tilde{\bL}^{1/2}\tilde{\bV}'$.
	
	\section{Review of Principle Component Analysis}
	PCA, among the most popular dimension reduction techniques, finds a new orthogonal basis for the data that maximizes the total variance in the projected space. To find new basis vectors $\be_1,\ldots,\be_p$, we sequentially solve the following constrained optimization:
	\begin{equation*}
		\begin{aligned}
			& \underset{\be_j}{ArgMax}
			& & Var(\bX \be_j) \\
			& \text{subject to:}
			& & \be_j'\be_j = 1,\\
			& 
			& & \be_j'\be_k = 0, \, j\neq k.
		\end{aligned}
	\end{equation*}
	Solving for $\be_1$ provides the principal direciton that captures the most variance. Given $\be_1$, $\be_2$ is the principle direction that captures the second-most variance while being orthogonal to $\be_1$; we continue in this manner until we solve for all $p$ basis vectors. The constraints ensure that we do not simply make $\be_j$ extremely large to achieve the maximization and also that the basis vectors are orthogonal. PCA has a simple, closed-form solution: $\be_j = \bv_j$, the eigenvector associated with the $j^{th}$ largest eigenvalue of $\bX'\bX$. We can then obtain orthogonal, high dimensional coordinates $\tilde{\bX}$ via the linear projection $\bX \bV$. To reduce the data to $m$ dimensions, we need only to keep the first $m$ columns of $\tilde{\bX}$. The quality of the projection can easily be quantified by the proportion of total variance preserved, given by  $(\sum_{j=1}^{m}\lambda_j)/(\sum_{j=1}^{p}\lambda_j)$. When the proportion of variance captured is higher, the projection more accurately reflects the high dimensional structure.
	
	\section{Review of Multidimensional Scaling}
	MDS is a general framework that creates low-dimensional projections that preserves high dimensional dissimilarities.  This is accomplished by minimizing a stress function. Many versions of stress functions exist \citep{kruskal1964multidimensional}, but one common choice  is the squared-loss between high and low dimensional dissimilarities.  That is, MDS finds low dimensional coordinates $\bz_i$ $(i=1,\ldots,n)$ my minimizing the stress function:
	\begin{align*}
		f(\bz_1,\ldots,\bz_n) &= \sum_{i=1}^{n} \sum_{j=1}^{n} \Big( \delta_{HD}(\bx_i,\bx_j) - \delta_{LD}(\bz_i,\bz_j) \Big)^2,
	\end{align*}
	where $\delta_{HD}$ and $\delta_{LD}$ are measures of dissimilarty between $\bx_i,\bx_j$ and $\bz_i,\bz_j$, respectively. By minimizing $f(\bz_1,\ldots,\bz_n)$, we obtain the optimal (in an average sense) projected coordinates, which we denote by $\hat{\bZ}=\underset{\bz_1,\ldots,\bz_n}{ArgMin}\,\,\,f(\bz_1,\ldots,\bz_n)$. Often, $\delta_{HD}$ and $\delta_{LD}$ are the same metric, but this is not necessary. Common choices include the Euclidean distance, Manhattan distance, squared-Euclidean dissimilarity, and cosine dissimilarity. Unlike PCA, MDS does not typically yield an analytical solution and usually requires numerical optimization. However, the flexibility in choosing the dissimilarities allows the analyst to specify what relationship to preseve in the projection. When the low-dimensional dissimilarity closely matches the high dimensional dissimilarity, there is low stress and the high dimensional relationship between the data is better preserved in the projection. 
	
	\subsection{Classical MDS, PCA, and MDS}
	\label{sec:classical_mds}
	The phrase ``Multidimensional Scaling'' is often ambiguous and used to refer to general dimension reduction. MDS is often thought to produce the same results as PCA when the Euclidean distance is used, but this is due to confusion with the nomenclature.  A description of Classical MDS \citep{torgerson1952multidimensional} proceeds.\\
	
	\noindent Given a matrix of pairwise Euclidean distances $\bD$ without knowing the raw data, find coordinates $\tilde{\bX}$ that preserve the distances. To accomplish this, Torgerson performs the following steps:
	
	\begin{enumerate}
		\item Square the pairwise distances $\bD$ to create $\bD^2$.
		\item Create a new matrix $\boldsymbol{B}$ by double centering $\bD^2$; that is, compute:
		\begin{align*}
			\boldsymbol{B} &= -\frac{1}{2}(\bI-\frac{1}{n}\bone \bone')\bD^2 (\bI-\frac{1}{n}\bone \bone').
		\end{align*}
		
		\item Denoting  $\lambda_1,\ldots,\lambda_p$ as the eigenvalues of $\boldsymbol{B}$ in descending order and $\bv_1,\ldots,\bv_p$ as the corresponding eigenvectors.  Let $\bV = (\bv_1,\ldots,\bv_p)$ and $\boldsymbol{\Lambda} = diag(\lambda_1,\ldots,\lambda_p)$.
		
		\item Create $\tilde{\bX} = \bV\boldsymbol{\Lambda}^{1/2}$, an $n \times p$ matrix preserving the distances in $\bD$. To create an $m<p$ dimensional projection, we use the first $m$ columns of $\tilde{\bX}$.
	\end{enumerate}
	The resulting solution $\tilde{\bX}$ is equivalent to projecting the original matrix $\bX$ linearly via its eigenvectors, which is exactly the result produced by PCA.
	
	MDS provides the user the flexibility to specify the dissimilarity measure to use; this results in the low dimensional projection that preserves the desired relationship. While Euclidean distances will not result in a projection equivalent to PCA, PCA is a particular case of MDS. If we define both the low- and high-dimensional dissimilarity metric to be the inner product, i.e. $\delta_{HD}(\bx_i,\bx_j)=\bx_i'\bx_j$ and $\delta_{LD}(\bz_i,\bz_j)=\bz_i'\bz_j$, MDS produces the same projection as PCA (Proof in Appendix \ref{sec:A1}).  Classical MDS (with Euclidean distances), PCA, and MDS (with inner-product dissimilarities) each create the same low dimensional projection.
	
	\subsection{Review of the PCA Biplot}
	
	Gabiel's PCA biplot \citep{gabriel1971biplot} is an extension of the PCA projection that labels the projection space in terms of the high dimensional attributes. Consider the SVD of the high dimensional data $\bX = \bU\bL^{1/2}\bV'$. $\bX$ can be futher decomposed into $b\bU\bL^{\alpha/2} \bL^{(1-\alpha)/2}\bV'/b$, where $\alpha \in [0,1]$ and $b$ is a scalar.  Gabriel shows that we can consider $b\bU\bL^{\alpha/2}$ as information about the observations and $\bV\bL^{(1-\alpha)/2}/b$ as information about the attributes embedded in the raw data.
	
	As in PCA, for dimension reduction we extract the first $m$ columns of each matrix, $\bU_1,\bL_1,$ and $\bV_1$. The matrix product $\tilde{\bX}=\bU_1\bL_1^{1/2}\bV_1'$ is a rank deficient approximation of $\bX$. To obtain a low dimensional projection of the observations, we plot the $n$ rows of $\bZ = b\bU_1\bL_1^{\alpha/2}$. Similarly, we plot the $p$ rows of $\bL_1^{(1-\alpha)/2}\bV_1/b$ as arrow-vectors (axes) from the origin, indicating the direction of the projection in terms of each attribute. Longer arrows represent the important variables driving the projection. 
	
\begin{figure}[H]
		\centering
		\includegraphics[width=5in]{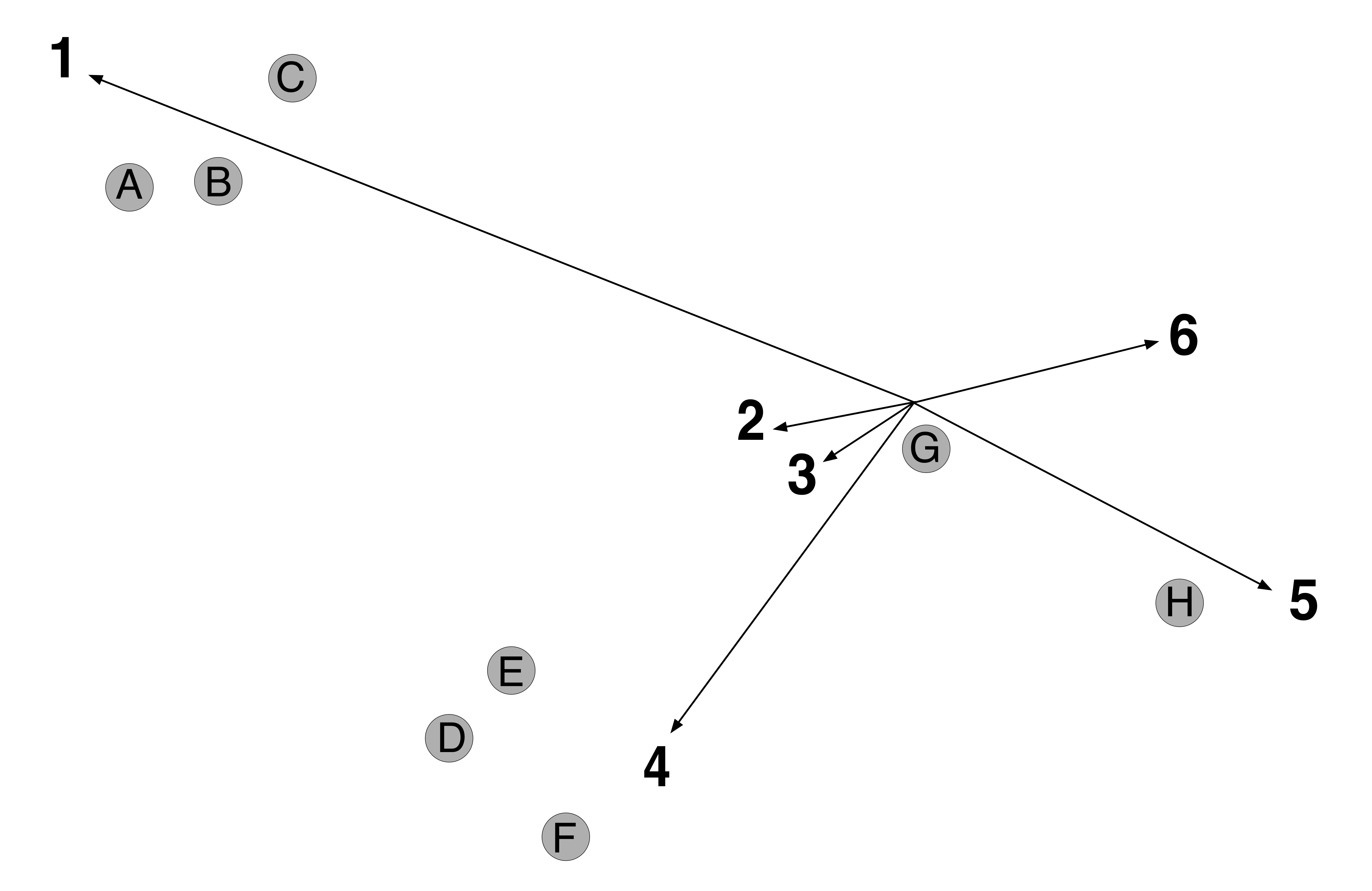}
		\caption{Example of a biplot with $6$ attributes. Observations A,B, and C strongly exhibit attribute $1$, while observations D, E, and F exhibit attribute $4$. Attribute $5$ is negatively correlated with $1$. Attributes $2,3,$ and $6$ explain less variability than the other attributes.} 
		\label{biplot_figure}
	\end{figure}

	The position of each projected observation, in relation to each attribute arrow, provides information about the orientation of the projection. For example, if an observation is far from the origin in the direction of a certain arrow, it strongly exhibits that attribute. Using this logic, we use the attributes to describe why certain observations are in close proximity (Figure \ref{biplot_figure}). 
	
	Since $\alpha$ and $b$ are continuous, there exist an uncountably infinite number of PCA biplot projections. Large values of $\alpha$ put more emphasis on maintaining the relationship between the observations and $b$ changes the scale of the plot. Gabriel recommends setting $b=n^{-1/2}$ so the lengths of the arrow will be close to the scale of $\bZ$. When we select $\alpha=1,b=1$, our low dimensional projection $\bZ=\tilde{\bU}\tilde{\bL}$ is equivalent to the PCA projection $\bZ=\bX\tilde{\bV}$. It follows that our low dimensional arrows are simply the rows of $\tilde{\bV}$. This is equivalent to projecting a high dimensional unit vector for each attribute in the direction of the eigenvectors of $\bX'\bX$. Clearly, the length of the axes for each attribute is left to the discretion of the user.  The axes are supposed to be representations of a high dimensional axis, which is technically of infinite length.  Gower \citep{gower2011understanding} provided an extension can add tick-marks on an infinite-length axis instead axes with finite-length. Since these choices for $\alpha$ and $b$ align with the idea of simply labeling the typical PCA projection, we continue their use for the remainder of the manuscript.
	
	\section{Biplots for MDS Projections}
	In this section, we develop biplots for any user-specified measures of dissimilarity. We use the stress function $f(\bz_1,\ldots,\bz_n)$ to not only match low-dimensional and high-dimensional dissimilarities between observations but also between the observations and each axis. We will approximate the continuous high dimensional axes by treating each as a finite sequence of uniformly spaced points.  For each point along the axis, we compute the high dimensional dissimilarity between the point and all of the observations. Then, using the low dimensional projection, we optimize the stress function to find the best low dimensional representation of the approximated axis. We repeat this process for each attribute to generate a complete set of low dimensional axes. 
	
	Let $\ba_{k,\ell} = (0,\ldots,\ell,\ldots,0)'$ denote a point along the $k^{th}$ attribute's high dimensional axis, $\ell$ units from the origin. We will find the corresponding low dimesional projection $\bb_{k,\ell}$ using the following procedure:
	
	\begin{enumerate}
		\item We optimize $f(\bz_1,\ldots,\bz_n)$ to obtain projections $\hat{\bz}_1,\ldots,\hat{\bz}_n$, which will remain fixed.
		
		\item For a uniformly spaced sequence $\ell \in \mathcal{L}=\{-c,\ldots,c\}$ for a constant $c$, find $$\underset{\bb_{k,\ell}}{ArgMin}\sum_{i=1}^{n}\big(\delta_{HD}(\bx_i,\ba_{k,\ell})-\delta_{LD}(\hat{\bz}_i,\bb_{k,\ell})\big)^2.$$
	\end{enumerate}
	
\begin{figure}[H]
		\centering
		\includegraphics[width=5in]{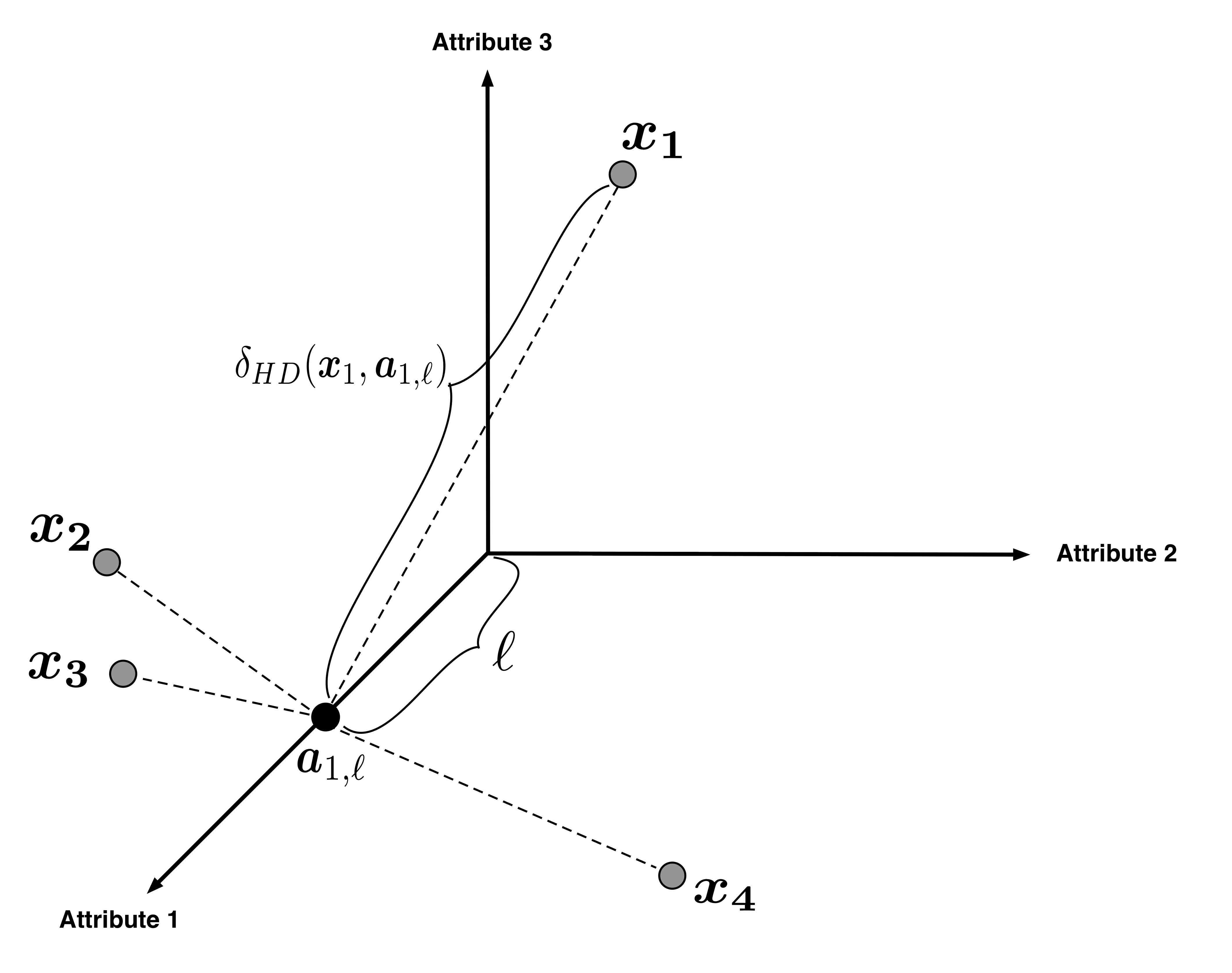}
		\caption{High dimensional visualization of the dissimilarities (Euclidean distance) between $\bx_1,\ldots,\bx_4$ and $\ba_{1,\ell}$.}
		\label{HD_figure}
	\end{figure}

\begin{figure}[H]
		\centering
		\includegraphics[width=5in]{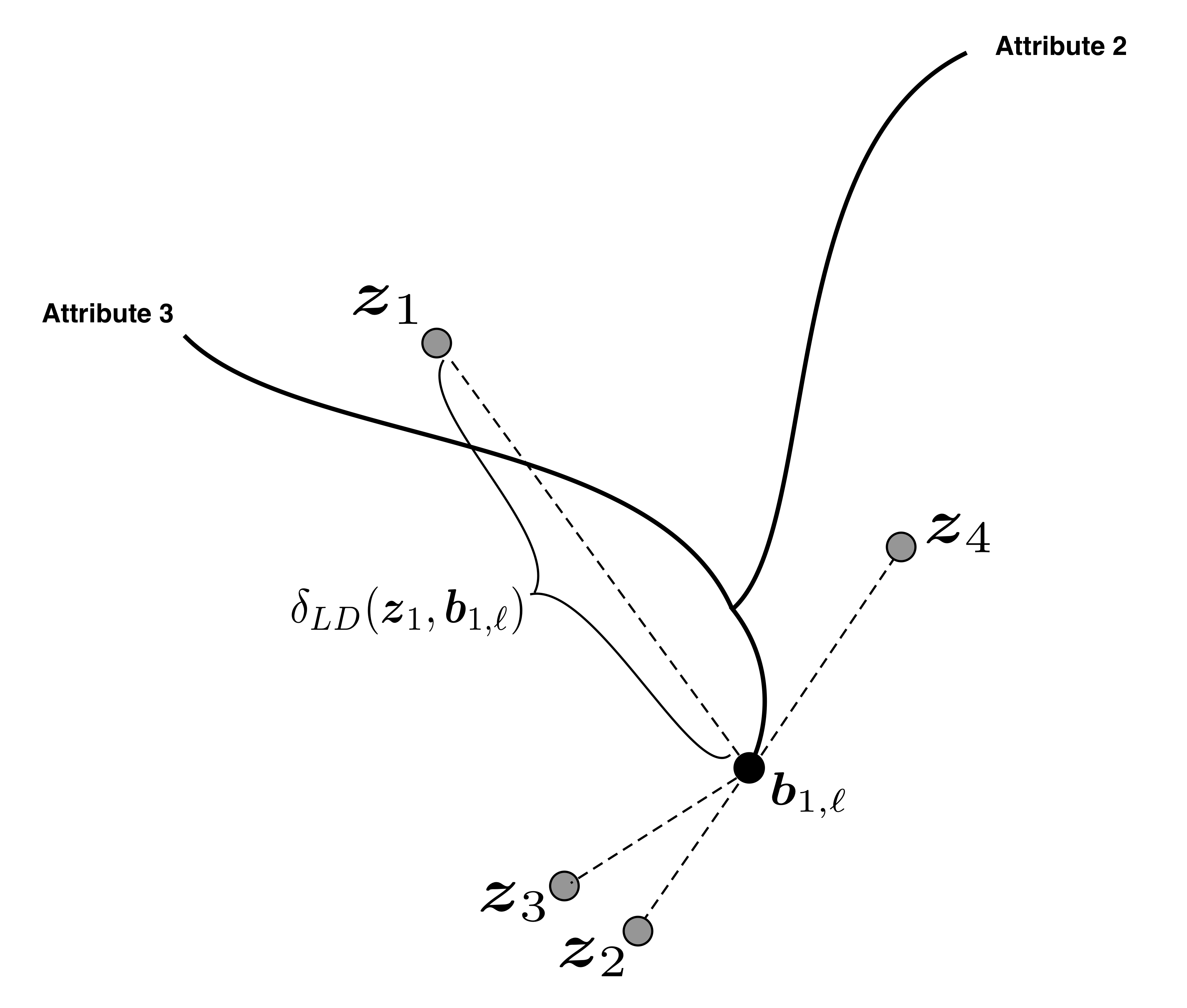}
		\caption{Low dimensional visualization of the dissimilarities (Euclidean distance) between $\bz_1,\ldots,\bz_4$ and $\bb_{1,\ell}$.} 
		\label{LD_figure}
	\end{figure}

	Fixing a particular $k$ and optimizing over the entire sequence for $\ell$, we obtain a full sequence of low dimensional points that we connect to form the axis for the $k^{th}$ attribute (Figures \ref{HD_figure} and \ref{LD_figure}).  We repeat step 2 for each attribute to solve for the entire set of axes.
	
	Since each LD axis is optimized independently, our Generalized MDS Biplot is easily parallelized. Then, for each $\ell$ the optimization simply solves for a $m\times1$ vector. Typically, the optimum for $\bb_{k,\ell}$ is very near $\bb_{k,\ell+\epsilon}$ for some small $\epsilon>0$, thus sequentially providing good initializations for the optimization along the entire axis.
	
	\subsection{High Stress Attributes}
\begin{figure}[H]
		\centering
		\includegraphics[width=5in]{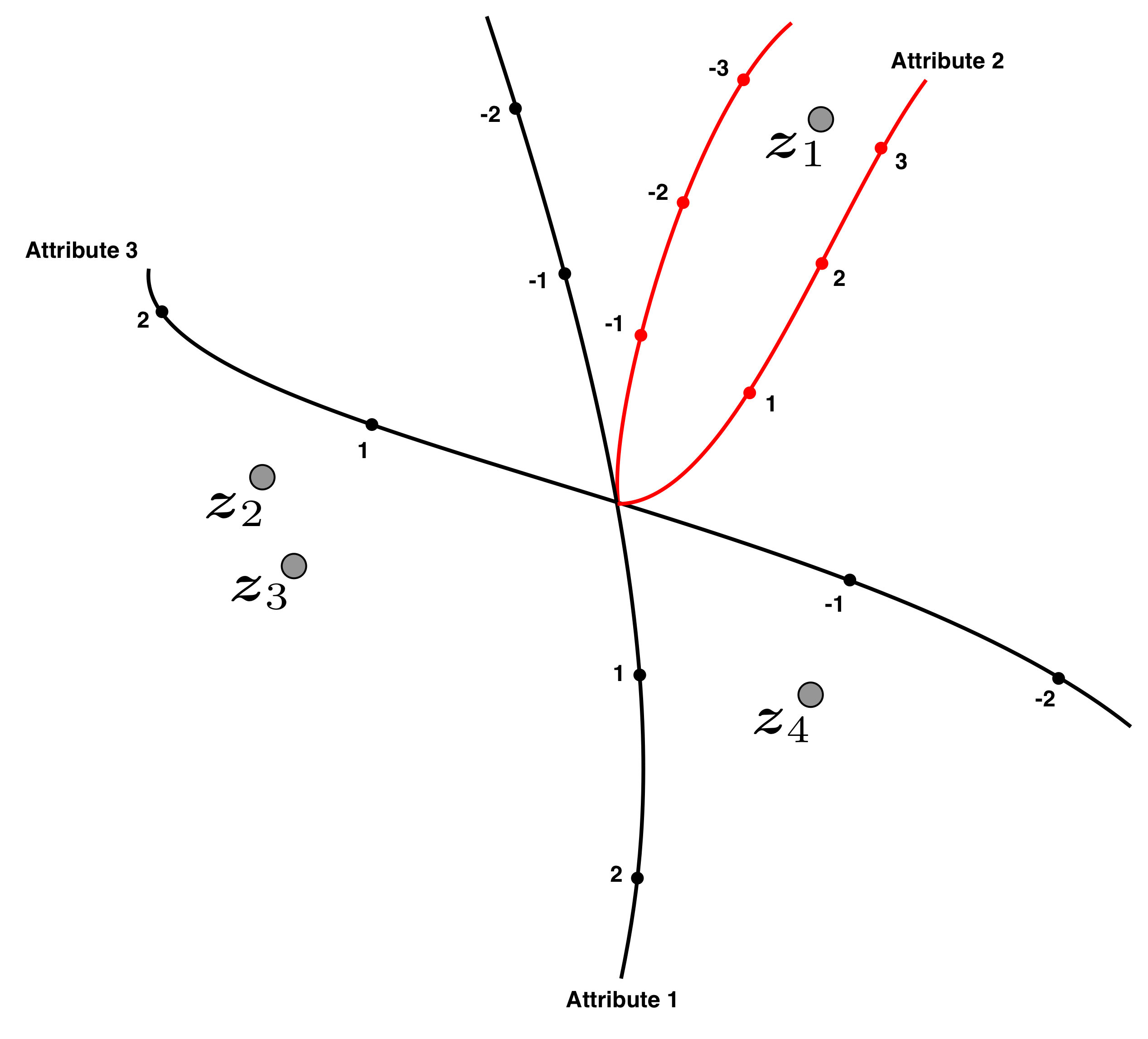}
		\caption{Generalized MDS Biplot for three attributes. The LD axis for attribute $2$ is U-shaped and $\bz_1$ is close to both the positive and negative sides of the axis.} 
		\label{U_axis}
	\end{figure}

	The Generalized MDS Biplots procedure will create LD axes for every attribute that is captured in the original MDS projection. For practical reasons, though, the user may not want to display some of the axes. In settings where $p$ is large, plotting all attributes would cover the entire projection, making it difficult to infer any structure. Additionally, since most dissimilarities induce nonlinear MDS projections, some LD axes may take on shapes that provide no benefit to the existing projection. For example, if an axis loops back towards itself, a neighboring observation would seem to be both be high and low in the attribute, simultaneously (Figure \ref{U_axis}). 
	
	We decide which axes to remove by considering the stress of the projection along each axis. Upon running both MDS and the Generalized MDS Biplot algorithms, we have $\hat{\bz}_1,\ldots,\hat{\bz}_n$ and $\hat{\bb}_{k,\ell}$ for $k \in \{1,\ldots,p\}$ and $\ell\in \mathcal{L}$. For one solution $\hat{\bb}_{k,\ell}$, we define 
	$$g(\hat{\bb}_{k,\ell}) = \sum_{i=1}^{n}\big(\delta_{HD}(\bx_i,\ba_{k,\ell})-\delta_{LD}(\hat{\bz}_i,\hat{\bb}_{k,\ell})\big)^2$$ to be the optimal stress resulting from the LD projection of $\ba_{k,\ell}$. Higher values of $g(\hat{\bb}_{k,\ell})$ correspond to HD axis points that do not fit as well in the existing MDS projection. To determine if an entire axis is highly stressed in the projection, we average over the axis. We denote the average stress for the projection of LD axis $k$ as 
	\begin{align*}
		G(k)&=\int_{-c}^{c} g(\hat{\bb}_{k,\ell})d\ell\\
		&\approx \frac{1}{\left\vert\mathcal{L}\right\vert}\sum_{\ell \in \mathcal{L}} g(\hat{\bb}_{k,\ell}).
	\end{align*}
	
	We recommend sequentially removing attributes from the Generalized MDS Biplot that have the highest values for $G(k)$ until a satisfactory projection is attained. The number of attributes to display is completely dependent on the application, but the aforementioned steps will generally remove the axes that are least useful in the projection. To demonstrate the concept, we perform a simulation study in the following section.
	
	\subsection{Simulation Study}
\begin{figure}[H]
		\centering
		\includegraphics[width=5in]{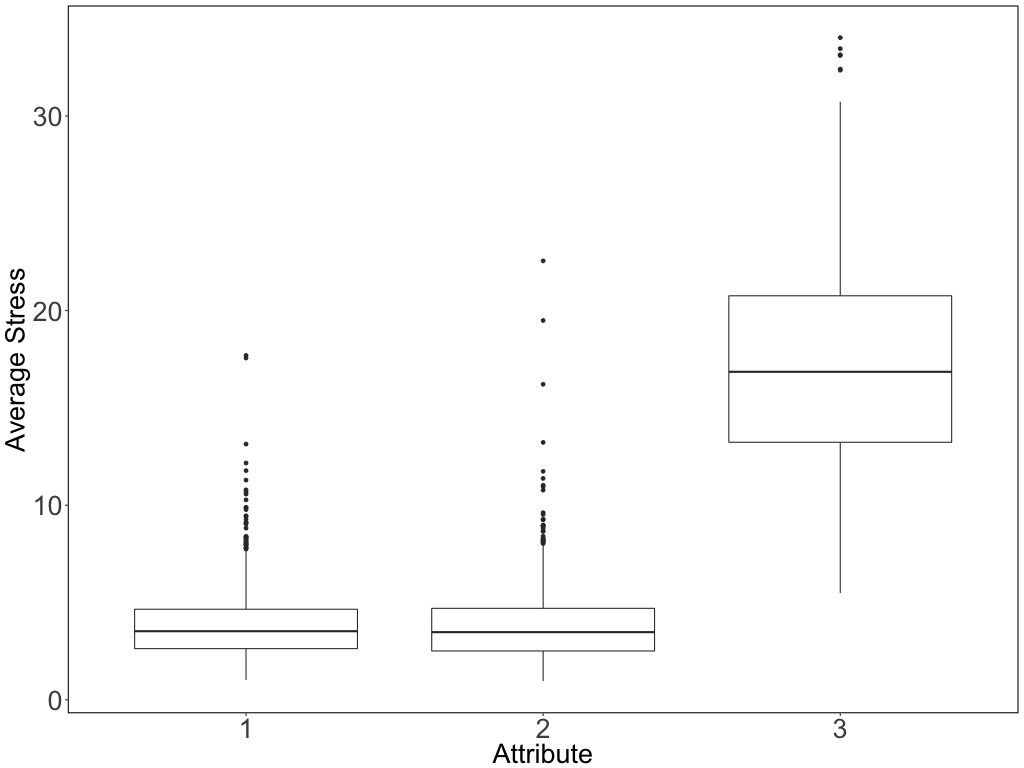}
		\caption{Average stress along each axis for $1,000$ simulations of the Generalized MDS Biplot using Manhattan distance as the HD dissimilarity.} 
		\label{sim_study}
	\end{figure}

	We simulate $n=25$ observations of $p=3$ attributes. Initially, the entire data matrix is simulated from a $Normal(\mu=0,\sigma=1)$. For each iteration of the simulation, we center all attributes to have mean $0$ and scale each attribute to have a different standard deviation. We simulate the standard deviations of attributes $1$ and $2$ from a $Uniform(0.5,1.0)$ and attribute $3$ from a $Uniform(0,0.5)$. As the variability of attribute $3$ approaches $0$, the HD data exists almost entirely on a $2$-dimensional surface and attribute $3$ contributes little to most HD dissimilarity choices (e.g. Euclidean, Manhattan, Cosine). We run MDS on the HD data and create the Generalized MDS Biplot using Manhattan distance as the HD dissimilarity. Finally, we compute the average stress, $G(k)$ for each $k$. The results of the simulation can be found in Figure \ref{sim_study}.
	
\begin{figure}[H]
		\centering
		\includegraphics[width=5in]{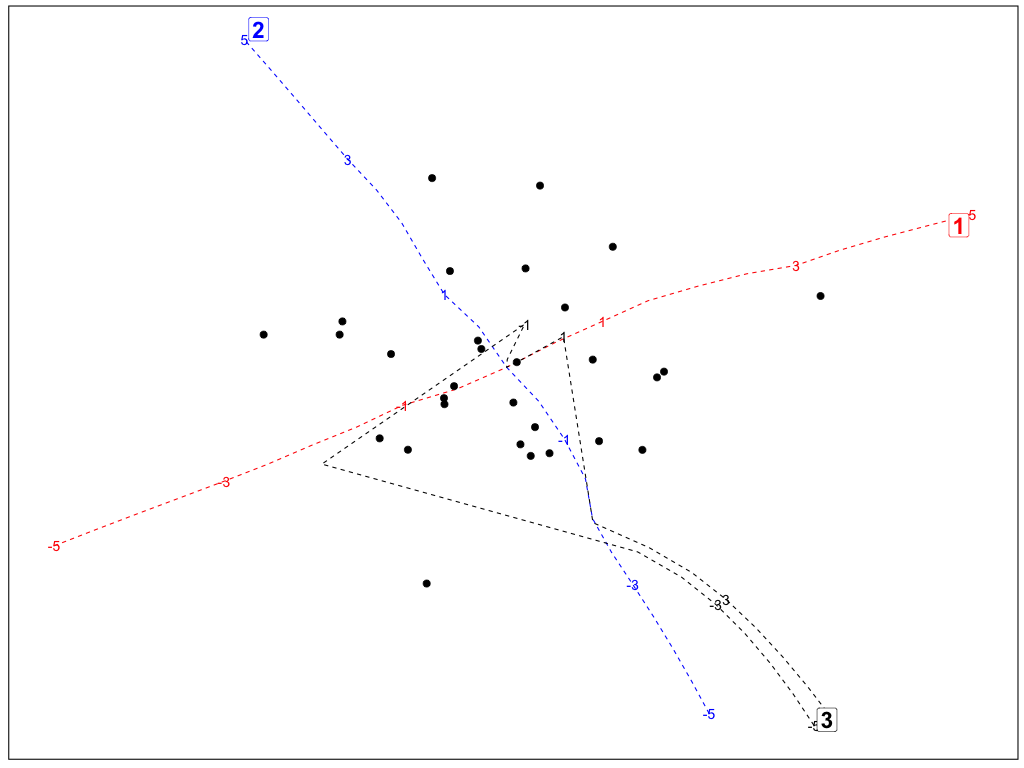}
		\caption{One example of the Generalized MDS Biplot from the simulation. The attribute with low variability is represented by an axis that loops back onto itself.} 
		\label{bad_axis_example}
	\end{figure}

	In almost every simulation, the average stress for attribute $3$ was higher than that of attributes $1$ and $2$. The stress discrepancy is greatest when the HD data is closer to being contained on the $2$-dimensional plane. Often, the simulations lead to Generalized MDS Biplots where attribute $3$ is represented by a U-shaped axis (Figure \ref{bad_axis_example}). The other axes in Figure \ref{bad_axis_example} remain quite straight and perpendicular since these are the attributes that define the plane containing the majority of the HD variability. Removing the axes with high average stress values provides a more useful and meaningful visualization to the user.
	
	\section{Case Study}
	\subsection{Data}
	Our dataset contains the $n=15$ Atlantic Coast Conference universities (Boston College, Clemson, Duke, Florida State, Georgia Tech, Louisville, Miami, North Carolina, North Carolina State, Notre Dame, Pittsburgh, Syracuse, Virginia, Virginia Tech, and Wake Forest), each with $p=8$ attributes: Student-to-Faculty ratio (Stud/Fac), Enrollment (Enroll), Percentage of students that are in a graduate program (GradStud), $75\%$-ile of ACT score (ACT), Percentage of applicants admitted (Admit), Graduation rate (GradRate), Percentage of students that are male (Male), and Average cost of attendance (AvgCost) \citep{nces}. To visualize these data, we apply three techniques: our Generalized MDS Biplot, Gower's Nonlinear Biplot, and Cheng and Mueller's Data Context Map. First, we briefly review the latter methods.\\
	
	\subsection{Nonlinear Biplot}
	The PCA Biplot gives context to the LD space by adding vectors, representing the attributes, to the existing LD projection of the observations. Gower generalizes the PCA biplot by permitting the use of any Euclidean embeddable dissimilarity $d_{ij}$. An $n \times n$ matrix $\bD$ with elements $d_{ij}$ is Euclidean embeddable if points $\tilde{\bx}_1,\ldots,\tilde{\bx}_n$ can be embedded in Euclidean space with Euclidean distance $\|\tilde{\bx}_i-\tilde{\bx}_j\|_2 = d_{ij} \, \forall i,j$ \citep{gower1986metric}. Examples of Euclidean embeddable dissimilarities include Euclidean distance $d_{ij}=\big(\sum_{k=1}^p(x_{ik}-x_{jk})^2\big)^{1/2}$, the square root of the Manhattan distance $d_{ij}=\big(\sum_{k=1}^p|x_{ik}-x_{jk}|\big)^{1/2}$, and Clark's distance $d_{ij}=\big(\sum_{k=1}^p\frac{x_{ik}-x_{jk}}{x_{ik}+x_{jk}}\big)^{1/2}$ (for nonnegative values) \citep{gower2005nonlinearity}. First, Gower uses classical MDS to create the low dimensional projection of the observations. Iteratively, each point along each high dimensional axis is treated as an $(n+1)^{st}$ point. Using the both the original high dimensional pairwise distances between observations and the distance from each observation to the $(n+1)^{st}$ point, Gower derives a closed-form linear mapping to obtain the low dimensional representation (details in Appendix \ref{sec:A2}).
	
	\subsection{Data Context Map}
	Cheng and Mueller simultaneously project the observations and single-point representations of each attribute. After scaling the high dimensional observations to the interval $\left[0,1\right]$, the attributes are treated as additional observations. They then create an $(n+p) \times (n+p)$ composite distance matrix (CDM) involving three different types of dissimilarities: between pairs of observations (DD), between pairs of attributes (VV), and between observations and attributes (DV). Choice of each dissimilarity is arbitrary, but the authors use the following:
	\begin{align*}
		\delta_{DD}(\bx_i,\bx_j) &= \Big(\sum_{k=1}^p(x_{ik}-x_{jk})^2\Big)^{1/2},\\
		\delta_{VV}(\bV_k,\bV_\ell) &= 1 - \rho(\bV_k,\bV_\ell),\\
		\delta_{DV}(\bx_i,\bV_k) &= 1 - x_{ik},
	\end{align*}
	where $\bV_k$ is the $k^{th}$ column (attribute) in the scaled data matrix and  $\rho(\bV_k,\bV_\ell)$ is the correlation between the $k^{th}$ and $\ell^{th}$ columns. Since each submatrix involves different dissimilarities, they are scaled to have the same mean before fusing them to create the CDM. Finally, the low dimensional projection is created by performing MDS on the CDM dissimilarities. 
	
	\subsection{Comparison of Methods}
	
	For the Generalized MDS Biplot (GMB), Nonlinear Biplot (NB), and Data Context Map (DCM), we will provide visualizations of the aforementioned ACC dataset. The various configurations across methods are described in Table \ref{configs}. Since the NB requires Euclidean embeddable distance functions, it cannot produce biplots for the Manhattan distance or Cosine dissimilarity. Unless noted otherwise, the LD dissimilarity will always be the Euclidean distance, arguably the most natural choice for two-dimensional visualization. For all GMB and NB projections, we center and scale the HD attributes to have mean $0$ and variance $1$. DCM requires each HD attribute to be scaled to the unit interval to be able to utilize the specified observation-to-attribute dissimilarity function. To produce each axis for the GMB and NB, we will use the HD axis sequence $\{-5.0,-4.9,\ldots,4.9,5.0\}$.\\
	
	\begin{table}[H]
		\centering
		\begin{tabular}{l|cccc}
			& Euclidean & Manhattan & Cosine \\ \hline \hline
			Nonlinear Biplot & X& N/A & N/A    \\
			Data Context Map & X & X & X  \\ 
			Generalized MDS Biplot & X & X & X     
		\end{tabular}
		\caption{HD dissimilarity configurations for each method. X denotes configurations that we visualize.}
		\label{configs}
	\end{table}
	
\begin{figure}[H]
		\centering
		\subfloat[NB (Euclidean), replicating the PCA Biplot\label{NL_Euclidean}]{%
			\includegraphics[width=0.45\textwidth]{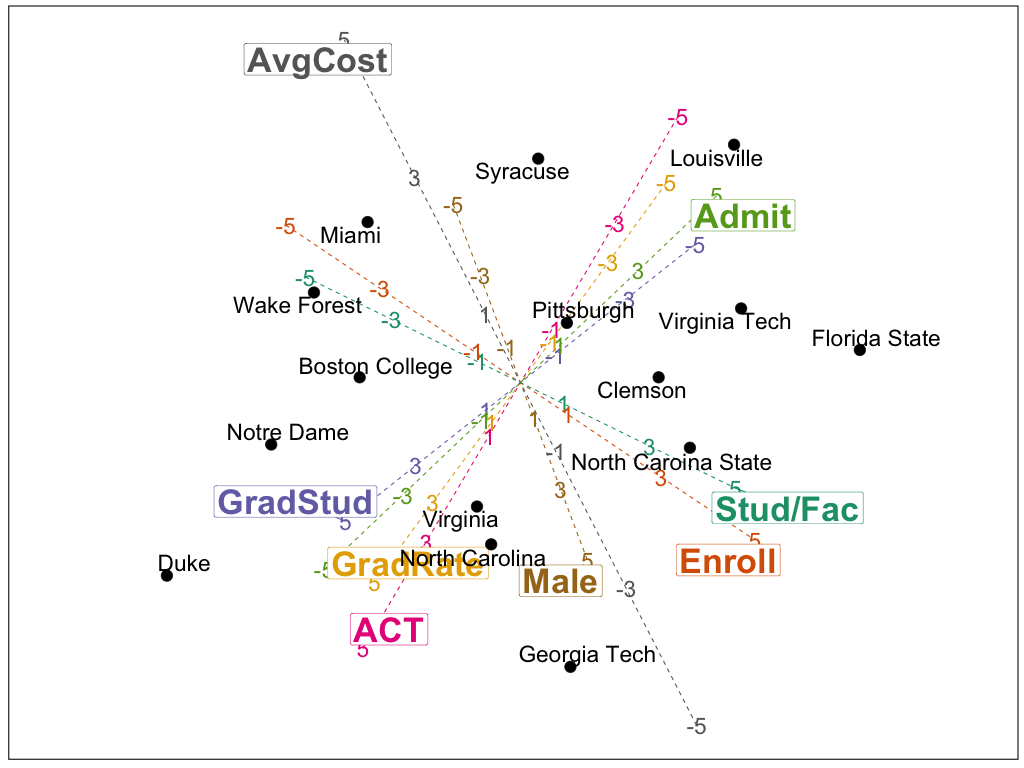}
		}
		\subfloat[DCM (Euclidean)\label{DCM_Euclidean}]{%
			\includegraphics[width=0.45\textwidth]{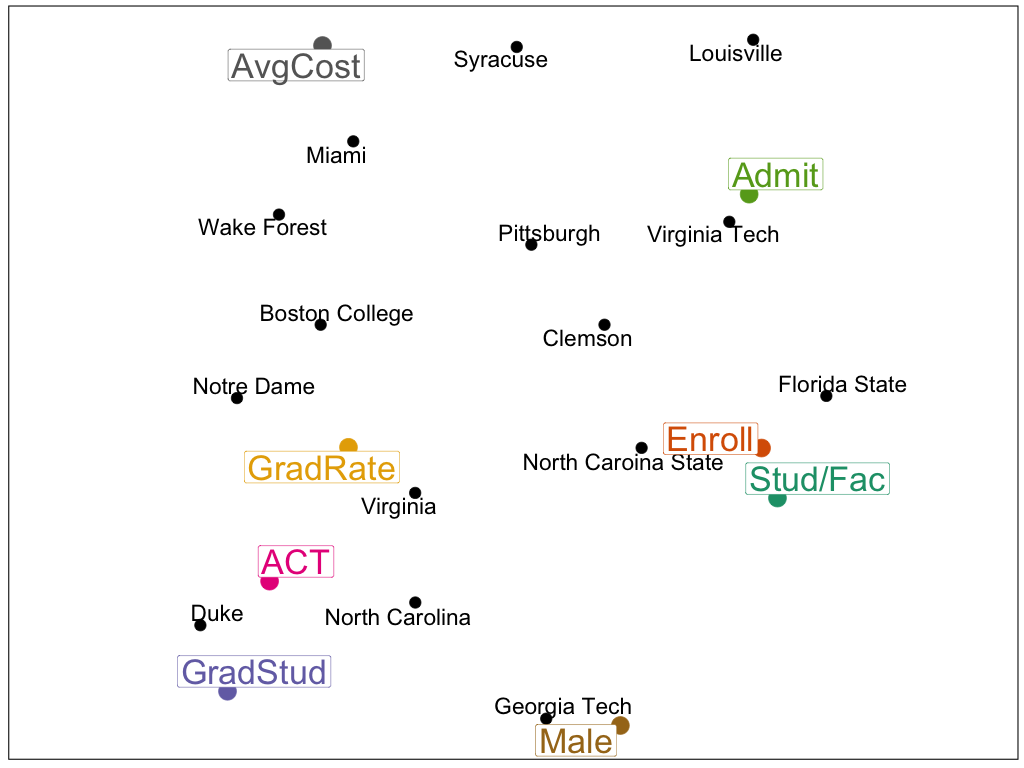}
		}
		\hspace{0mm}
		\centering
		\subfloat[GMB (Euclidean) \label{GMB_Euclidean}]{%
			\includegraphics[width=0.45\textwidth]{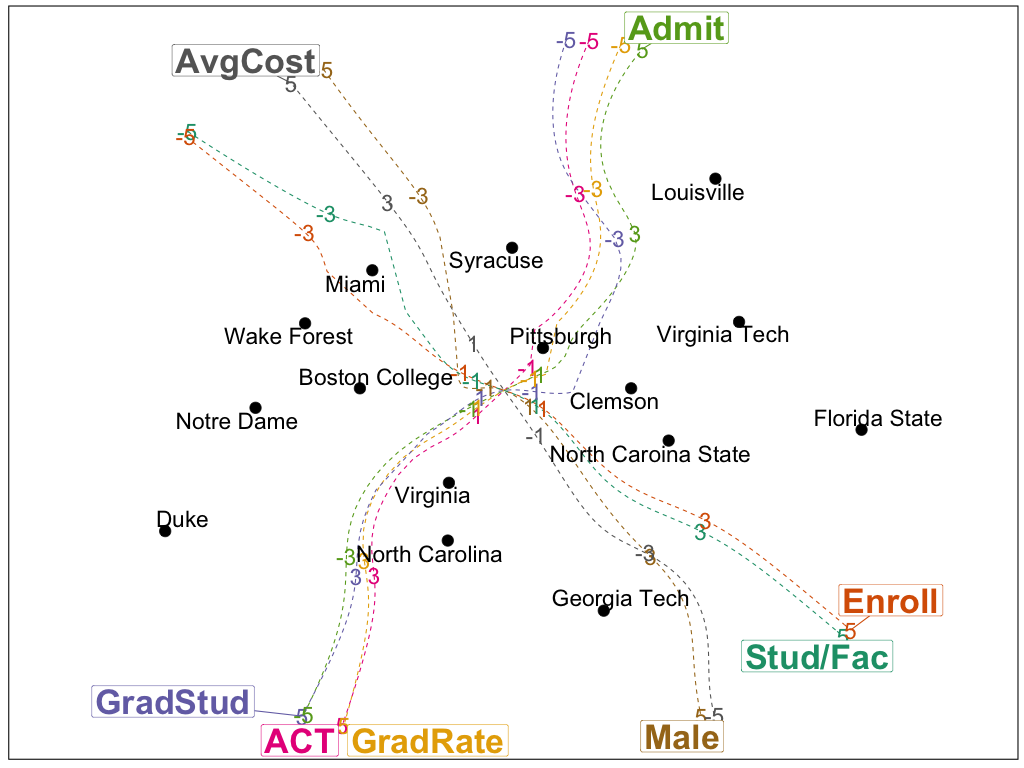}
		}
		\subfloat[GMB (Inner-product for both HD and LD), replicating both the PCA Biplot and NB with Euclidean distance) \label{GMB_IP}]{%
			\includegraphics[width=0.45\textwidth]{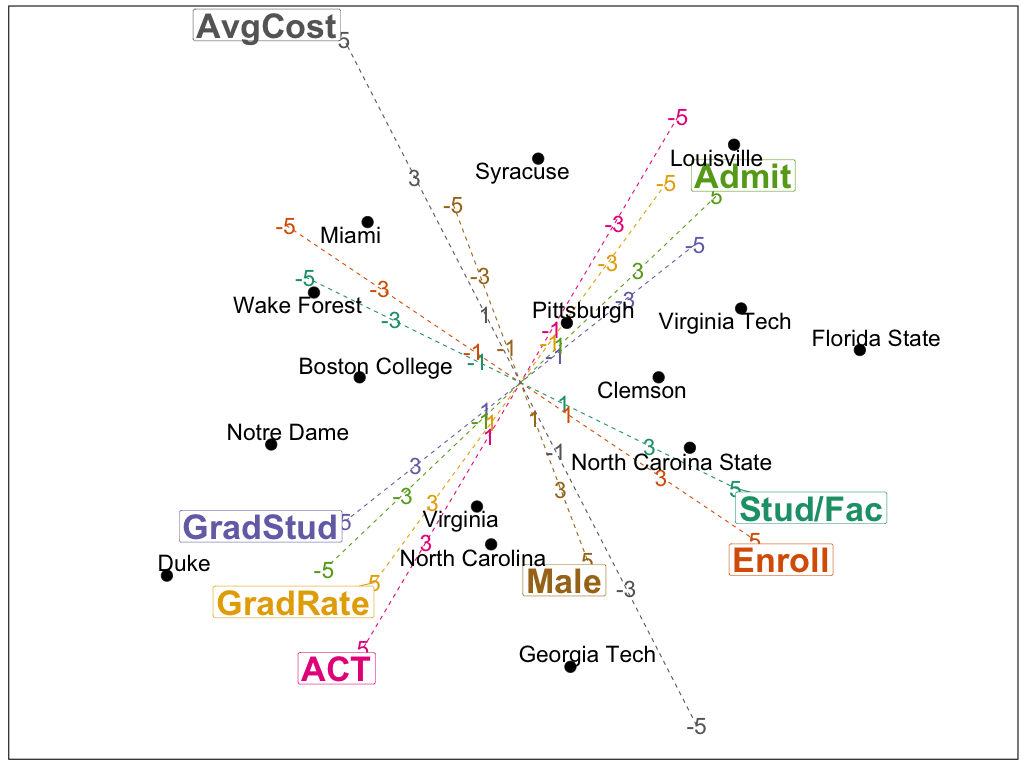}
		}
		\caption{Comparison using Euclidean distance as HD dissimilarity.}
	\end{figure}

	\noindent \underline{\textbf{Euclidean distance}}: The NB with Euclidean distance (Figure \ref{NL_Euclidean}) exactly matches the projected observations and axes of the PCA Biplot, a result that follows from the use of Classical MDS (as detailed in Section \ref{sec:classical_mds}) to produce the projection. That is, when we choose to use the Euclidean distance, the NB is actually a linear projection of the HD data and HD axes.
	
	The DCM treats the attributes as additional observations. As a result, each attribute is represented by a single point, rather than an axis (Figure \ref{DCM_Euclidean}). Since the projection of the attributes and observations occur simultaneously, observations are often strongly drawn towards the attribute they strongly exhibit. For example, Duke has the highest percentage of graduate students and Georgia Tech has the highest proportion of students that are male. For any observation that has the highest value of an attribute, the HD observation-to-attribute dissimilarity will always be $0$. When $p$ is large, relative to $n$, this effect is more prominent.
	
	Within our GMB framework, the HD Euclidean distance does not change linearly as $\ell$ varies (Figure \ref{GMB_Euclidean}). Consequently, the LD axes tend to curve, often with each axis gravitating towards observations that strongly exhibit the attribute. Unlike the DCM, the axis itself will not impact the projection of the observations; it only provides labels to the already existing projected space. While the LD axes follow the same general trajectory as the PCA Biplot, we can actually reproduce it precisely. When $\delta_{HD}$ and $\delta_{LD}$ are both chosen to be inner-products, the GMB exactly replicates the PCA Biplot (Figure \ref{GMB_IP}). A proof of this result can be found in Appendix \ref{sec:A3}.
	
\begin{figure}[H]
		
		\centering
		\subfloat[DCM (Manhattan)\label{DCM_Manhattan}]{%
			\includegraphics[width=0.45\textwidth]{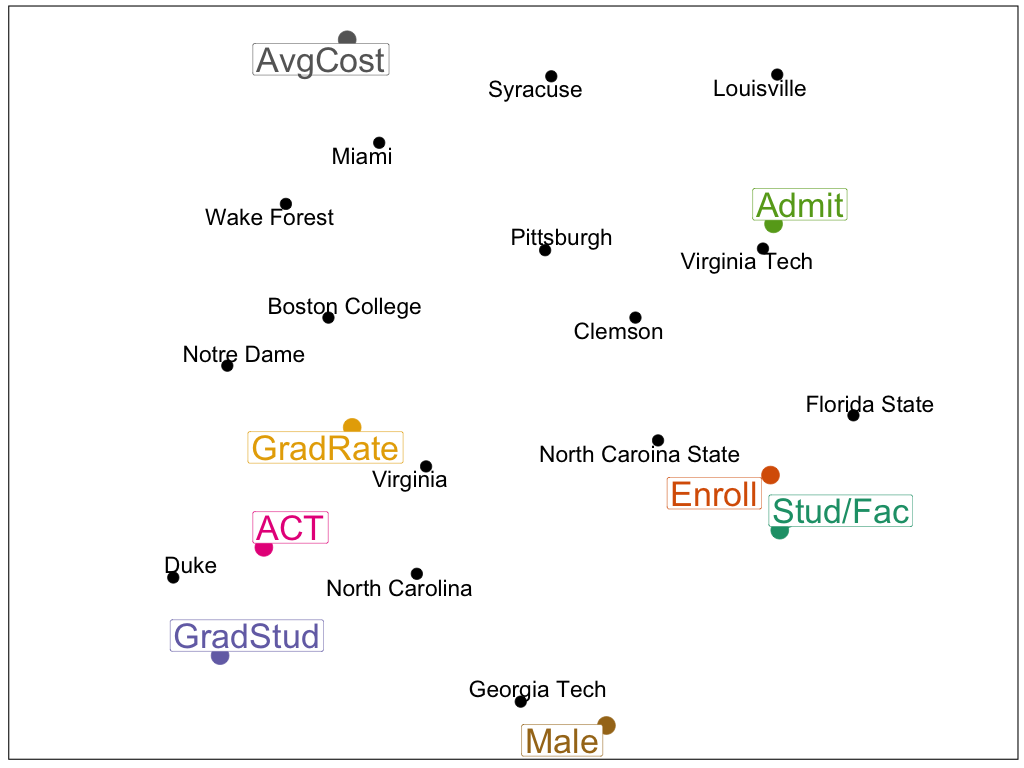}
		}
		\subfloat[GMB (Manhattan)\label{GMB_manhattan}]{%
			\includegraphics[width=0.45\textwidth]{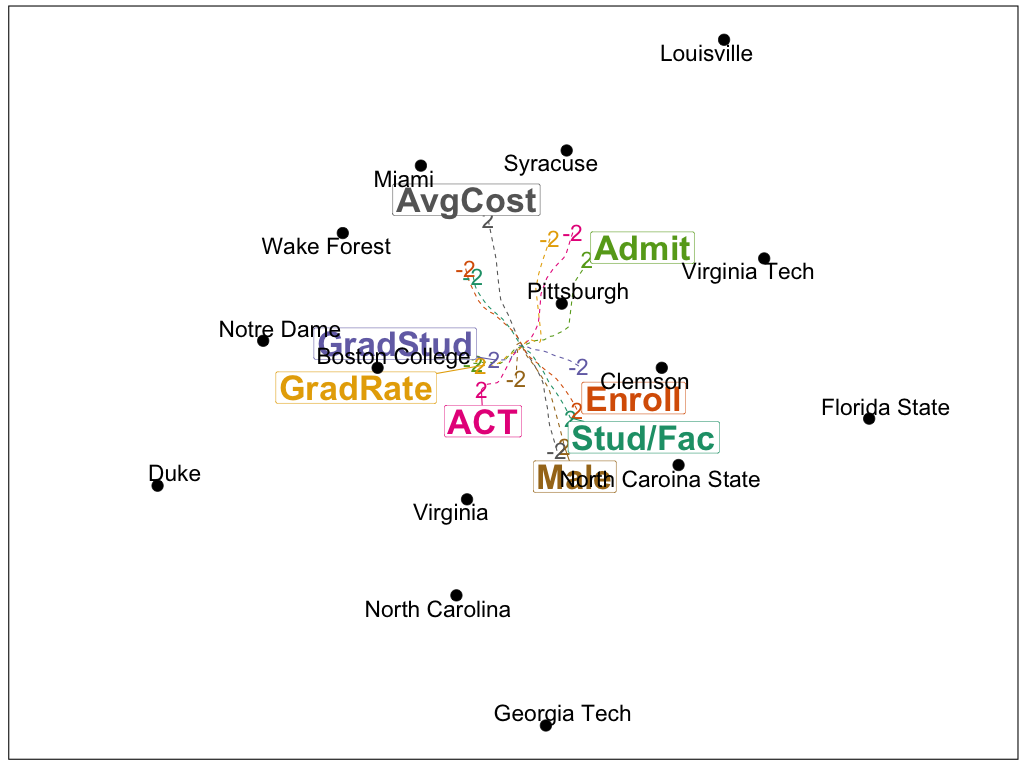}
		}
		\caption{Comparison using the Manhattan distance as HD dissimilarity.}
		
	\end{figure}

	\noindent \underline{\textbf{Manhattan distance}}: The GMB with Manhattan distance produces a projection for the observations that looks similar to the GMB with Euclidean distance (Figure \ref{GMB_manhattan}). We do observe different behavior with the axes, though. We elect to only plot the LD axes for $\ell \in \lbrack -2,2 \rbrack$. First, almost all of the standardized HD data is between these values. When we expand the axis range to values beyond the range of our HD data, several of GMB axes sharply turn towards the same region in the top/center of the projection. However, using our selected range, we observe that the overall orientation of the axes is similar to both the Euclidean distance GMB and the PCA biplot.
	
	The DMC with Manhattan distance (Figure \ref{DCM_Manhattan}) is very similar to the DMC with Euclidean distance (Figure \ref{DCM_Euclidean}). The effect of the observation-to-attribute dissimilarity largely overpowers the choice of Manhattan distance for the observation-to-observation dissimilarity. MDS projections using Manhattan distance tend to produce right-angles and diamond shapes, but this is not the case for the DCM. Similar behavior is examined in the next section for the Cosine dissimilarity.
	
\begin{figure}[H]
		\centering
		\subfloat[DCM (Cosine)\label{DCM_Cosine}]{%
			\includegraphics[width=0.45\textwidth]{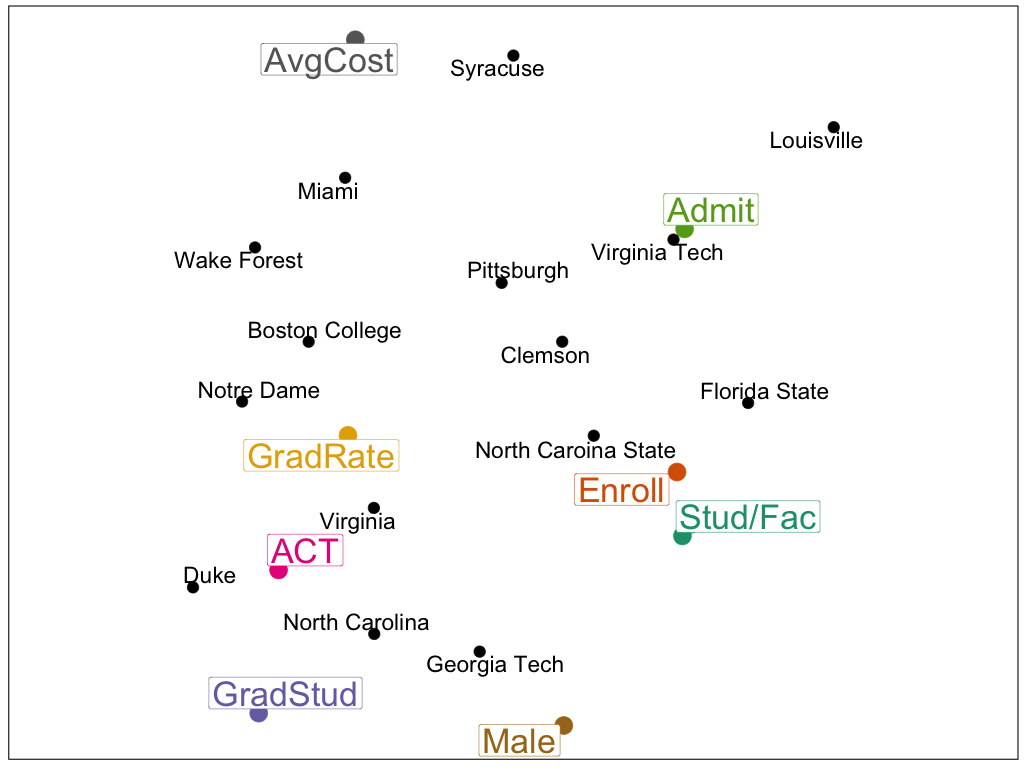}
		}
		\hspace{0mm}
		\centering    
		\subfloat[GMB (Cosine) \label{GMB_cosine}]{%
			\includegraphics[width=0.45\textwidth]{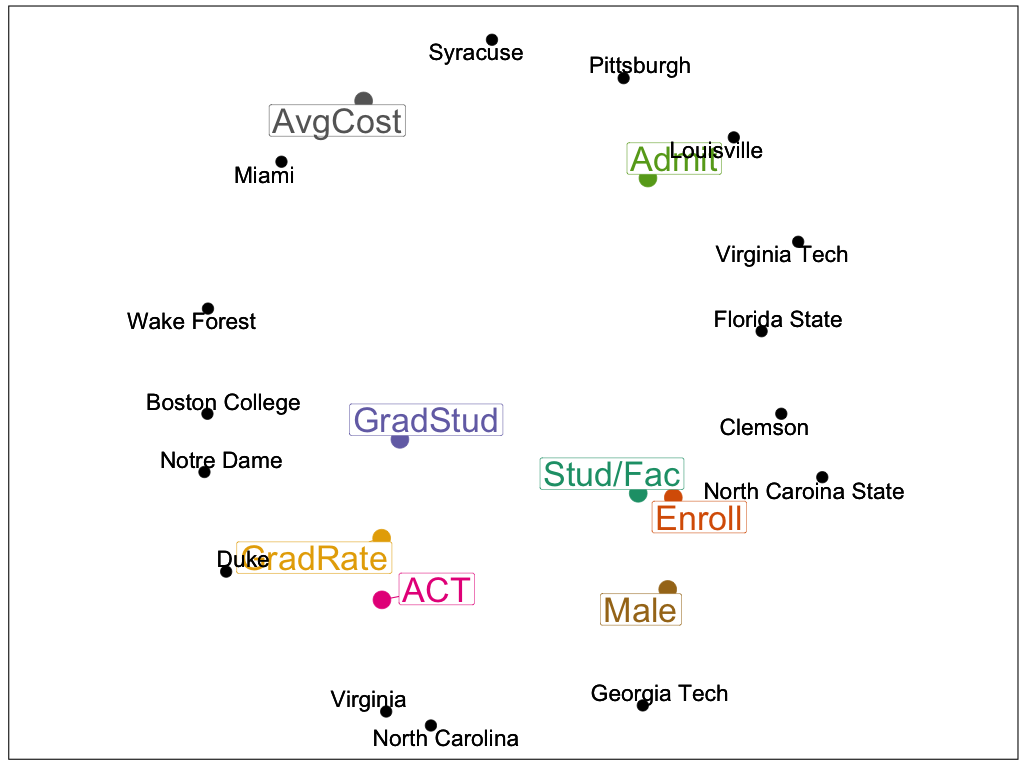}
		}
		\caption{Comparison using the Cosine dissimilarity as HD dissimilarity.}
	\end{figure}

	\noindent \underline{\textbf{Cosine dissimilarity}}: The Cosine dissimilarity is a scaled version of the inner-product and a measurement of the angle between two vectors. Consequently, $\delta_{HD}(\bx_i,\ba_{k,\ell}) = \delta_{HD}(\bx_i,\ba_{k,\ell'})$ for all $\ell$ and $\ell'$. The optimization along all HD axis points for a given attribute will result in the same LD axis point, providing single-point LD axes (Figure \ref{GMB_cosine}). In this way, the Cosine dissimilarity provides a layout most similar to the DCM. Just as Manhattan distance MDS projections tend to be diamond-like, the Cosine dissimilarity tends to produce circular projections. Within the DCM, the interaction of the attributes with the observation prevents the circular projection (Figure \ref{DCM_Cosine}). Again, the GMB is only labeling the existing MDS projection in accordance with the stress function used for the MDS projection.
	
	\section{Conclusion}
	\label{sec:conc}
	Data visualization is a useful tool that allows analysts to more easily understand the structure HD data. While PCA and MDS produce an LD visualization of the observations, users lose a sense of how the HD attributes are involved in the projection. When the projection is created via PCA, the PCA Biplot provides meaning, in terms of the attributes, of the LD space. The Non-linear Biplot extends the PCA Biplot to any Euclidean embeddable distance function, but many measures of dissimilarity are not Euclidean embeddable. The Data Context Map treats attributes as observations and simultaneously projects both into LD space, but doing so distorts the projection of the observations. The treatment of each attribute as an observation also increases the computational time needed to create the projection. Our Generalized MDS Biplot labels an already existing MDS projection without changing the observation-to-observation relationship. By treating each axis as an independent sequence of HD points, our algorithm is easily parallelizable and creates a meaningful projection with only trivial optimizations.
	
	\appendix
	\section{Appendix}
	
	\subsection{Proof of MDS and PCA equivalence under the inner-product dissimilarity}
	\label{sec:A1}
	Let $\delta_{HD}(\bx_i,\bx_j)=\bx_i'\bx_j$ and $\delta_{LD}(\bz_i,\bz_j)=\bz_i'\bz_j$.
	\begin{align*}
		\underset{\bz_1,\ldots,\bz_n}{ArgMin} \,\,\, f(\bz_1,\ldots,\bz_n) &= \underset{\bz_1,\ldots,\bz_n}{ArgMin} \,\,\, \sum_{i=1}^{n} \sum_{j=1}^{n} \Big( \bx_i'\bx_j - \bz_i'\bz_j \Big)^2\\
		&= \underset{\bz_1,\ldots,\bz_n}{ArgMin} \,\,\, Trace\Big( (\bX-\bZ)'(\bX-\bZ) \Big)\\
		&= \underset{\bz_1,\ldots,\bz_n}{ArgMin} \,\,\, \|\bX \bX' - \bZ \bZ'\|_F,
	\end{align*}
	where F denotes the Frobenius norm. Utilizing the SVD for both $\bX$ and $\bZ$, we write $\bX \bX'=\bU\bL\bU'$, $\bZ \bZ' = \tilde{\bU}\tilde{\bL}\tilde{\bU}'$, and the  function as:
	\begin{align*}
		f(\bz_1,\ldots,\bz_n) &=  \| \bU \bL \bU' - \tilde{\bU} \tilde{\bL} \tilde{\bU}' \|_F.
	\end{align*}
	Using the Eckart-Young-Mirsky theorem \citep{eckart1936approximation}, this expression is minimized when $\tilde{\bU}$ is the first $m$ columns of $\bU$ and $\tilde{\bL}$ is a diagonal matrix with the $m$ largest eigenvalues from $\bL$.  It follows that:
	\begin{align*}
		\bZ \bZ' &= \tilde{\bU}\tilde{\bL}\tilde{\bU}'\\
		&= \bU_1 \bL_1 \bU_1'\\
		&= (\bU_1 \bL_1^{1/2})(\bU_1 \bL_1^{1/2})'.
	\end{align*}
	The normalized eigenvectors of $\bX\bX'$ can easily be converted to normalized eigenvectors of $\bX'\bX$ by the relationship, $\bU_1=\bX\bV_1\bL_1^{-1/2}$.
	\begin{align*}
		\bZ &= (\bU_1 \bL_1^{1/2})\\
		&= \bX\bV_1\bL_1^{-1/2}\bL_1^{1/2}\\
		&= \bX \bV_1 ,
	\end{align*}
	which is the exact projection produced by PCA. Therefore, the PCA projection minimizes the MDS stress function when both the high and low dimensional dissimilarity metrics are defined to be the inner-product. 
	
	\subsection{Formulae for Gower's nonlinear biplot}
	\label{sec:A2}
	Let $\bx_{n+1}$ denote an $(n+1)^{th}$ point along a high dimensional axis. To obtain its low dimensional projection, calculate the following steps:
	
	\begin{enumerate}
		\item Calculate low dimensional coordinates $\bZ$ for $\bx_1,\ldots\bx_n$ via classical MDS.
		
		\item Compute the high dimensional distance between the $(n+1)^{st}$ point and the existing $n$ observations, $d_{i,n+1}$ for $i=1,\ldots,n$.
		
		\item Define an $n \times 1$ vector $\boldsymbol{d}$ with elements:
		$$ \frac{1}{n}\sum_{j=1}^{n}d_{ij}^2 - \frac{1}{2n^2}\sum_{i=1}^{n}\sum_{j=1}^{n}d_{ij}^2 - (d_{i,n+1})^2. $$
		
		\item Project $\boldsymbol{d}$ as the low dimensional axis $\boldsymbol{y}_{n+1}$ by the projection:
		$$\boldsymbol{y}_{n+1}=\frac{1}{2}(\bZ'\bZ)^{-1}\bZ'\boldsymbol{d}.$$
	\end{enumerate}
	
	\subsection{Proof of generalized MDS biplot and PCA biplot equivalence under the inner product dissimilarity}
	\label{sec:A3}
	Let $\delta_{HD}(\bx_i,\bx_j)=\bx_i'\bx_j$ and $\delta_{LD}(\bz_i,\bz_j)=\bz_i'\bz_j$.   Therefore, our low dimensional projection of $\bX$ is $\bZ=\bX\bV_1=\bU_1\bL_1$. Let $\ba_{k,\ell} = (0,\ldots,\ell,\ldots,0)'$ denote a point along the axis of the $k^{th}$ attribute of length $\ell$. We solve the following optimization:
	\begin{alignat*}{3}
		\underset{\bb_{k,\ell}}{ArgMin} \, f(\bb_{k,\ell}) &= \underset{\bb_{k,\ell}}{ArgMin} \sum_{i=1}^{n}\big(\bx_i'\ba_{k,\ell}-\bz_i'\bb_{k,\ell}\big)^2 \\
		&= \underset{\bb_{k,\ell}}{ArgMin} \big(\bX\ba_{k,\ell} -\bZ\bb_{k,\ell} \big)'\big(\bX\ba_{k,\ell} -\bZ\bb_{k,\ell} \big)
	\end{alignat*}
	To minimize the stress, we differentiate with respect to $\bb_{k,\ell}$.
	\begin{alignat*}{3}
		f(\bb_{k,\ell}) &= \big(\bX\ba_k -\bZ\bb_{k,\ell} \big)'\big(\bX\ba_k -\bZ\bb_{k,\ell} \big)\\
		\frac{\partial f(\bb_{k,\ell})}{\partial \bb_{k,\ell}} &= -2(\bZ'\bX) \ba_{k,\ell} + 2(\bZ'\bZ)\bb_{k,\ell}
	\end{alignat*}
	Setting the system of derivatives equal to $\b0$, we solve for the solution, $\hat{\bb_{k,\ell}}$.
	\begin{align*}
		(\bZ'\bZ) \hat{\bb}_{k,\ell} &= (\bZ'\bX)\ba_{k,\ell}\\
		\bL_1 \hat{\bb}_{k,\ell} &= (\bL_1^{1/2} \bU_1' \bU \bL^{1/2} \bV')\ba_{k,\ell}\\
		\hat{\bb}_{k,\ell}&= \bV_1' (0,\ldots,\ell,\ldots,0)'\\
		\hat{\bb}_{k,\ell} &= \ell (v_{k1},\ldots,v_{km})',
	\end{align*}
	where $(v_{k1},\ldots,v_{km})'$ denotes the $k^{th}$ row of the matrix $\bV_1$. When $\ell=1$, the result exactly matches the PCA biplot, which projects the unit vector in the direction of the first $m$ eigenvectors. It is trivial to show that this solution is indeed a maximum.  The Hessian is
	\begin{align*}
		\frac{\partial^2 f(\bb_k)}{\partial\bb_k^2} &= 2\bL_1.
	\end{align*}
	Since all the eigenvalues of $\bX'\bX$ are positive, $\bL_1$ is positive definite and the solution is a maximum.

	\RaggedRight
	\bibliographystyle{agsm}
	\bibliography{biplotref}
	
\end{document}